\documentclass[12pt,a4paper]{article}

\usepackage{amsmath,amssymb,amsthm}
\usepackage{cite}

\usepackage[dvipdfmx]{graphicx}
\usepackage[english]{babel} 

\usepackage[charter]{mathdesign}

\newlength{\topdummy}
\begin{document}

\begin{titlepage}

\begin{flushright}
UT--14--20
\end{flushright}

\vskip 2.7cm
\begin{center}

{\huge \bf
Lepton Universality Test of Extra Leptons \\ \vspace*{2mm}
using Hadron Decay
}

\vskip .95in

{\large
\ Motoi Endo and Takahiro Yoshinaga}
\vskip 0.4in

{\large
{\it 
Department of Physics,  University of Tokyo, Tokyo 113-0033, Japan
}}

\end{center}
\vskip .95in

\begin{abstract}

We study the flavor universality of the lepton couplings with the $W$ and $Z$ bosons. 
Currently, it is constrained by the electroweak precision observables.
We focus on the leptonic decay of the pion.
It is shown that the decay is expected to have a comparable or better sensitivity to the lepton-flavor universality in the near future. 
Furthermore, it provides a complementary information in determining the $W\ell\nu$ couplings without assuming new physics models.
This result is applied to a model of the extra lepton.

\end{abstract}

\end{titlepage}

\section{Introduction}

Physics beyond the standard model (SM) has been studied extensively in many years. 
One of the simplest extensions of SM is to introduce exotic leptons in addition to the SM matters.
They are assumed to have the same quantum numbers under the SM gauge symmetries as the SM leptons. 
In general, they have vector-like masses and couple to the SM leptons via Yukawa interactions. 

Such extra leptons are often motivated by the excess of the experimental result of the anomalous magnetic moment of the muon (muon $g-2$)~\cite{g-2_bnl2010} against the SM prediction~\cite{g-2_hagiwara2011,g-2_davier2010}.
The discrepancy between the experimental value and its SM prediction is more than 3$\sigma $ levels as $a_{\mu }^{{\rm exp}} - a_{\mu }^{{\rm SM}} = (26.1 \pm 8.0) \cdot 10^{-10}$~\cite{g-2_hagiwara2011} or $(28.7 \pm 8.0) \cdot 10^{-10}$~\cite{g-2_davier2010}, depending on the estimation of the leading hadronic contributions to the theoretical value. 
This excess may suggest the existence of new physics in the electroweak scale.
A solution is provided by the models in which the muon Yukawa interaction is generated through the mixing with the extra leptons~\cite{Kannike:2011ng,Dermisek:2013gta}.
The 3$\sigma $ discrepancy is solved when they are relatively light.

The extra leptons are searched for through direct productions at colliders or measurements of the precision observables. 
In particular, they can affect the interactions among the SM particles.
When the extra leptons have vector-like masses and couple to the SM leptons, they generally contribute to the SM lepton couplings with the $W$, $Z$, and Higgs bosons (see e.g., Ref.~\cite{Kannike:2011ng}).
These contributions depend on flavors. 
Those to the electron couplings have been constrained by the electroweak precision observables (EWPOs)~\cite{ALEPH:2005ab}. 
On the other hand, the muon interactions may include sizable corrections, if the extra leptons solve the muon $g-2$ anomaly (see Refs.~\cite{Kannike:2011ng,Dermisek:2013gta}). 
Such a flavor non-universality is realized by the Yukawa interactions between the extra leptons and the SM ones. 
Thus, the effects of the extra leptons may be probed as non-universalities of the SM lepton interactions.

In this letter, we investigate the precision observables that are sensitive to the lepton non-universality. 
In particular, we focus on the flavor universality between the electron and muon interactions with the $W$ and $Z$ bosons. 
Currently, they are constrained most severely by EWPOs~\cite{delAguila:2008pw}. 
In this letter, we discuss the leptonic pion decay, $\pi \to \ell \nu _{\ell} $, which is also sensitive to the lepton universality~\cite{Kumar:2013qya}. 
Its measurement as well as the SM prediction is very accurate and expected to be improved in the near future. 
It will be shown that the pion decay can provide a better probe of the non-universality than EWPOs. 
Furthermore, it provides a complementary information in determining the lepton couplings with the $W$ bosons. 

This letter is organized as follows.
In the first part of this letter, corrections to the lepton couplings with the $W$ and $Z$ bosons are studied in a model-independent way.
The leptonic pion decay is sensitive to the flavor universality of the $W\ell\nu$ couplings as shown in Sec.~\ref{sec:pion}. 
We discuss current status and future prospects of the corrections to the $W\ell\nu$ couplings. 
On the other hand, the lepton non-universality is already limited by EWPOs.
In particular, they contribute to the theoretical value of the Fermi constant, $G_F$, the $W$ decay width, and the $Z$ couplings with the leptons. 
The first two observables are sensitive to the corrections to the $W\ell\nu$ couplings, while the last one restricts those to the $Z\ell\ell$ interactions. 
The $Z\ell\ell$ coupling is naively correlated with that of the $W$ boson.
However, the relation  depends on details of the extra lepton models (see e.g., Ref.~\cite{Kannike:2011ng}). 
Thus, the contributions to the $W$ and $Z$ bosons are discussed seprately in Sec.~\ref{sec:ewpo}. 
We show the result of searching for the electron/muon flavor universality in Sec.~\ref{sec:result}.
It is applied to a model of the extra leptons in Sec.~\ref{sec:model}.
The last section is devoted to the conclusion.


\section{Pion Decay}\label{sec:pion} 

First of all, let us define the correction to the $W\ell\nu$ coupling.
The interaction among the $W$ boson, the left-handed charged lepton and the neutrino is
\begin{equation}
 \mathcal{L} = \frac{g}{\sqrt{2}} \Delta g^{W \ell \nu} \,
 \bar{\ell }_L \gamma ^{\mu } \nu _{\ell }\, W^+ _{\mu } + {\rm h.c.},  
 \label{eq:lagW}
\end{equation}
where $g$ is the SU(2)$_L$ gauge coupling, and $\ell = e$, $\mu $.
The coefficient $\Delta g^{W \ell \nu}$ is decomposed into the SM value and its correction as
\begin{equation}
 \Delta g^{We\nu _e} = 1 - \delta g^{We\nu _e}, ~~~~ 
 \Delta g^{W \mu \nu  _{\mu }} = 1 - \delta g^{W \mu \nu  _{\mu }}, 
 \label{eq:corrW}
\end{equation}
where the unity in the right-hand side corresponds to the SM prediction.
The lepton universality is preserved when $\Delta g^{We\nu _e} = \Delta g^{W \mu \nu  _{\mu }}$.
It is realized in SM due to the hypothesis that the leptons have identical gauge charges and their chiralities mix only via the Higgs interactions. 
It can be violated by the new physics contributions.
They are represented by $\delta g^{W \ell \nu}$ $(\ll 1)$. 

One of the most sensitive probes of the electron--muon flavor universality is provided by the leptonic decay of the pion, $\pi \to \ell \nu _{\ell }$~\cite{Bryman:2011zz,Kumar:2013qya}.
The decay proceeds via the $W$ boson exchange, and thus, the rate is sensitive to $\delta g^{W \ell \nu}$.
A ratio of the electric and muonic decays, $\pi \to e \nu _e / \mu \nu _{\mu }$, is represented as (cf.~Refs.~\cite{Kinoshita:1959ha,Marciano:1976jc,Cirigliano:2007xi})
\begin{align}
  R_{\pi \to e/\mu } &= 
  \frac{\Gamma (\pi \to e \nu_e (\gamma ))}{\Gamma (\pi \to \mu \nu_{\mu } (\gamma ))} 
  = \left| \frac{\Delta g^{We\nu _e}}{\Delta g^{W \mu \nu  _{\mu }} } \right| ^2 
  \frac{m_e ^2 }{m_{\mu }^2} 
  \left( \frac{1- m_e ^2/m_{\pi} ^2}{1- m_{\mu }^2/m_{\pi } ^2} \right) ^2 
  (1+\delta R_{\pi \to e/ \mu } ) 
  \notag \\
  & \simeq 
  (1- 2\delta g^{We\nu _e} + 2 \delta g^{W \mu \nu  _{\mu }}  ) 
  R_{\pi \to e/\mu }^{{\rm SM}}, 
  \label{eq:deltapi}
\end{align}
where $R_{\pi \to e/\mu }^{{\rm SM}}$ is the SM prediction. 
Although radiative corrections $\delta R_{\pi \to e/ \mu }$ could also depend on $\Delta g^{W \ell \nu}$, the effect is sufficiently small and neglected here. 
It is stressed that the ratio is not polluted by hadronic uncertainties.
It is calculated with a high accuracy as~\cite{Bryman:2011zz,Kumar:2013qya,Kinoshita:1959ha,Marciano:1976jc,Cirigliano:2007xi}\footnote
{
In Ref.~\cite{Cirigliano:2007ga}, two-loop contributions of $\mathcal{O} (e^2 p^4)$ are calculated within the chiral perturbation theory. 
The SM prediction becomes $(1.2352 \pm 0.0001) \times 10^{-4}$, which is more accurate than Eq.~\eqref{eq:decaypi1}.
Here, we adopt a conservative result.
Anyway, the accuracy of the theoretical prediction is much better than that of the experimental result. 
} 
\begin{equation}
 R_{\pi \to e/\mu }^{{\rm SM}} = (1.2351 \pm 0.0002) \times 10^{-4}.
 \label{eq:decaypi1}
\end{equation}
Thus, the uncertainty of the SM prediction is 0.02\%. 

On the other hand, the experimental result of the leptonic pion decay has been reported as~\cite{Beringer:1900zz}
\begin{equation}
 R_{\pi \to e/\mu }^{{\rm exp}} = (1.230 \pm 0.004) \times 10^{-4},
 \label{eq:decaypi2}
\end{equation}
where the results of TRIUMF~\cite{Britton:1992pg} and PSI~\cite{Czapek:1993kc} are averaged by PDG~\cite{Beringer:1900zz}.
The precision of the current experimental result is 0.3\%.
It is larger by an order of magnitude than the theoretical prediction.
There are ongoing experiments that can improve the experimental precision. 
Two experiments are in progress at TRIUMF~\cite{AguilarArevalo:2010fv} and PSI~\cite{Pocanic:2003pf}, which are expected to achieve an accuracy of 0.05--0.06\% or better for $R_{\pi \to e/\mu }$. 
Also, the PEN experiment~\cite{PEN} have accumulated $>10^7$ events of $\pi \to e\nu$ and $>10^8$ events of $\pi \to \mu \to e$ during 2008--2010. 
The analysis of those data is under way to determine $R_{\pi \to e/\mu }$.
The precision goal is about 0.05\%~\cite{Kumar:2013qya}, which is comparable to TRIUMF and PSI. 

From the theoretical prediction and the experimental result, the correction to the $W\ell\nu$ coupling is limited in the range:
\begin{equation}
 \delta g^{W \mu \nu  _{\mu }} - \delta g^{We\nu _e} = (-2.0 \pm 1.6) \times 10^{-3}.
 \label{eq:senst1}
\end{equation}
Since the experimental result is consistent with the theoretical prediction, the correction must be suppressed. 
If the uncertainty of the experimental value can be reduced to be 0.05\% in future, the correction is determined as
\begin{equation}
 \delta g^{W \mu \nu  _{\mu }} - \delta g^{We\nu _e} = (-2.0 \pm 0.3) \times 10^{-3},
 \label{eq:senst2}
\end{equation}
where the central value is fixed to be that in Eq.~\eqref{eq:senst1}.
Thus, it is found that the sensitivity of the correction to the $W\ell\nu$ coupling is expected to be improved by a factor of 5 or more.
It is also stressed that the pion decay determines $\delta g^{W \mu \nu  _{\mu }}$ {\it minus} $\delta g^{We\nu _e}$.
The relative sign comes from Eq.~\eqref{eq:deltapi}, where the ratio of the decay channels is considered. 

Let us comment on the $K$ and $\tau$ decays.
Their leptonic decays are also sensitive to the lepton universality.
Current experimental results of the $K$ decay have reached a precision of 0.4\% at KLOE~\cite{Ambrosino:2009aa} and NA62~\cite{Goudzovski:2010uk}. 
The sensitivity is expected to achieve 0.1\% in the near future~\cite{Goudzovski:2010uk}.
The SM prediction is estimated very accurately~\cite{Cirigliano:2007xi}.
The correction to the $W\ell\nu$ coupling is limited as $\delta g^{W \mu \nu  _{\mu }} - \delta g^{We\nu _e} = (2.2 \pm 2.0) \times 10^{-3}$ (cf., \cite{Pich:2013lsa}).
The accuracy is expected to become about $5\times 10^{-4}$ in future.
It is worse by a factor of 2 than Eq.~\eqref{eq:senst2}.
On the other hand, the current precision of the $W\ell\nu$ coupling from the $\tau$ decay is $\delta g^{W \mu \nu  _{\mu }} - \delta g^{We\nu _e} = (-1.8 \pm 1.4) \times 10^{-3}$ \cite{Pich:2013lsa}.
This accuracy is better than that of Eq.~\eqref{eq:senst1}.
The future sensitivity may be expected to be $5\times 10^{-4}$, according to the estimation for the SuperB project~\cite{Bona:2007qt}.
Since the pion decay is most sensitive to the lepton universality in future, we focus on it in this letter.

\section{Electroweak Precision}\label{sec:ewpo}

Measurements of the $Z$ boson decays have been used to test SM and constrain the new physics.
This is often called the electroweak precision test (EWPT).
In particular, contributions to the SM lepton couplings with the $W$ and $Z$ bosons are constrained. 
When the lepton couplings are modified, there are mainly three types of the contributions to EWPOs. 
The couplings to the $Z$ boson are constrained by the measurements of the $Z$ boson decaying to leptons. 
Those to the $W$ boson contribute to the determination of the Fermi constant $G_F$ through 
the measurements of the lifetime of the muon. 
Also, they change the decay width of the $W$ boson.
In this section, we study these observables to restrict the lepton flavor universality.

Let us first focus on the $W\ell\nu$ coupling, which is directly correlated with the leptonic pion decay.
Since the Fermi constant $G_F$ is determined by the lifetime of the muon, it is affected by corrections to the $W e \nu _e$ and $W \mu \nu _{\mu }$ couplings. 
The contribution to $G_F$ is represented as
\begin{equation}
 G_F = G^{\text{SM }}_F +  \frac{g^2}{4\sqrt{2} m_W^2} \Delta \bar{\delta } _G, 
\end{equation}
where $G^{\text{SM}}_F$ is the SM prediction.
Effects of the new physics are denoted by $\Delta \bar{\delta } _G$. 
When there are corrections to the $W\ell\nu$ coupling \eqref{eq:corrW}, it becomes
\begin{align}
 \Delta \bar{\delta } _G = -( \delta g^{We\nu _e} +  \delta g^{W \mu \nu  _{\mu }}) .
 \label{eq:corrGF}
\end{align}
It is noticed that $\delta g^{W \mu \nu  _{\mu }}$ has the same sign as $\delta g^{We\nu _e}$. 
This is contrasted to the decays of the pion (or $K$, $\tau$), as shown in Eq.~\eqref{eq:deltapi}. 
The correction to the Fermi constant affects the $Zff$ couplings, and the mass and decay width of the $W$ boson. 

The corrections to the $W\ell\nu$ coupling also contributes to the decay width of the $W$ boson, $\Gamma_W$, through the decay channel, $W \to \ell\nu_\ell$.
The correction to $\Gamma_W$ is represented as 
\begin{equation}
 \Gamma_W = \Gamma_W^{\text{SM }} + \frac{G_F m_W^3}{6\sqrt{2}\pi} \Delta \bar\delta_{\Gamma W}, 
\end{equation}
where $\Gamma_W^{\text{SM }}$ is the SM prediction.
The corrections to the $W\ell\nu$ coupling \eqref{eq:corrW} induce $\Delta \bar\delta_{\Gamma W}$ as
\begin{align}
 \Delta \bar\delta_{\Gamma W} = - 2( \delta g^{We\nu _e} +  \delta g^{W \mu \nu  _{\mu }}) .
 \label{eq:corrGW}
\end{align}
We checked that this contribution is subdominant in EWPT compared to Eq.~\eqref{eq:corrGF}. 
 
Next, let us consider the interaction between the $Z$ boson and the charged leptons.
The effective $Zff$ coupling, $g^{Z\ell\ell}_{L, R}$, is defined as
\begin{equation}
  \mathcal{L} = 
  \frac{e}{s_W c_W} 
  \bar{\ell } 
  \left[ g^{Z\ell\ell}_L \gamma ^{\mu } P_L + g^{Z\ell\ell}_R \gamma ^{\mu } P_R \right]
  \ell\, Z_{\mu } + {\rm h.c.},
\end{equation}
where $e = \sqrt{4\pi\alpha}$, $s_W = \sin \theta _W$ and $c_W = \cos \theta _W$ with the Weinberg angle $\theta _W$, and $P_{L(R)}$ is the left- (right-) handed projection operator.
The $Z$ boson decay observables are affected by the new physics through these effective couplings as well as $G_F$.

In this letter, we focus on corrections to the left-handed interaction, $g^{Z\ell\ell}_L$.
It is decomposed into the SM value and its correction as
\begin{equation}
 g^{Zee}_L = \left.g^{Zee}_L\right|_{\rm SM} + \delta g^{Zee}_L, ~~~~
 g^{Z\mu\mu}_L = \left.g^{Z\mu\mu}_L\right|_{\rm SM} + \delta g^{Z\mu\mu}_L,
 \label{eq:corrZLL}
\end{equation} 
where $\left.g^{Z\ell\ell}_L\right|_{\rm SM}$ is the SM prediction. 
The new physics contribution $\delta g^{Z\ell\ell}_L$ is related to the $W\ell\nu$ coupling \eqref{eq:corrW}.
However, details depend on new physics models (see e.g., Ref.~\cite{Kannike:2011ng}). 
On the other hand, $\delta g^{Z\ell\ell}_R$ is independent of the left-handed coupling.
For instance, in the extra lepton model which will be discussed later, $\delta g^{W \ell \nu} =  \delta g^{Z\ell\ell}_L$ is approximately satisfied, while the contributions to $\delta g^{Z\ell\ell}_R$ can be suppressed. 

The lepton flavor universality is studied by EWPT via the above contributions. In this letter, the following observables are considered:
\begin{align}
&\text{line-shape and FB asymmetry : }
 \Gamma _Z,~\sigma _h^0, ~R_f, ~A^f_{\text{FB}} ~(f= e, \mu , \tau ), \notag \\
& \tau ~\text{polarization : }
A_{\tau },~A_e, \notag \\
& b ~\text{and}~ c ~\text{quarks : }
R_f, ~A^f_{\text{FB}}, A_f ~(f=b, c), \notag \\
&\text{SLD results : }
A_f ~(f= e,\mu ,\tau) , \notag \\
& W~\text{mass and width : }
m_W, ~\Gamma _W. \label{eq:obs}
\end{align}
The experimental results and their correlation coefficients are found in Ref.~\cite{ALEPH:2005ab}, while the measurements of the mass and width of the $W$ boson are updated by Tevatron~\cite{Group:2012gb}.
Since the correlation coefficients of the $W$ boson observables are not available, we simply assume that they are independent of each other and the other observables.

The experimental values are compared with the theoretical predictions. 
The latter includes the contributions from SM and the new physics.
Because of the high precision of the experimental values, it is important to include higher-order corrections in the SM calculations.
Recently, the full EW two-loop contributions of the closed fermion loops are completed~\cite{Awramik:2003rn, Awramik:2006uz, Awramik:2008gi, Freitas:2013dpa, Freitas:2014hra}. 
We use the fitting formulae of EWPOs obtained in Ref.~\cite{Freitas:2014hra}, which also include other radiative corrections. 
On the other hand, the new physics contributes to EWPOs through Eqs.~\eqref{eq:corrGF}, \eqref{eq:corrGW} and \eqref{eq:corrZLL}. 
Eqs.~\eqref{eq:corrGF} and \eqref{eq:corrZLL} are taken into account by using the fitting formulae explored in Ref.~\cite{Cho:1999km, Cho:2011rk}, while Eq.~\eqref{eq:corrGW} is included by modifying them. 
In the theoretical calculations, the following SM inputs are used~\cite{Beringer:1900zz,g-2_davier2010}: 
\begin{align}
&m_t = 173.07 \pm 0.52 \pm 0.72, \\
&m_H = 125.9 \pm 0.4, \\
&\alpha _s(m_Z) = 0.1185 \pm 0.0006, \\
&\Delta \alpha ^{(5)}_{\text{had}}(m_Z) = 0.02757 \pm 0.00010.
\end{align}
In addition, the $Z$ boson mass $m_Z$, the Fermi constant $G_F$ and the fine structure constant $\alpha $ are $m_Z = 91.1876 \pm 0.0021$, $G_F = 1.1663787(6) \times 10^{-5}$ and $\alpha = 1/137.035999074$, respectively \cite{Beringer:1900zz}.
In the analysis, we take $m_Z$, $G_F$ and $\alpha$ as constant values, since their uncertainties are sufficiently small. 
In order to limit the lepton non-universality, we perform the global $\chi ^2$ fit of the above 21 observables.
The result is shown in the next section.


\section{Result}\label{sec:result}

\begin{figure}[htbp]
\begin{center}
 \includegraphics[width=8cm]{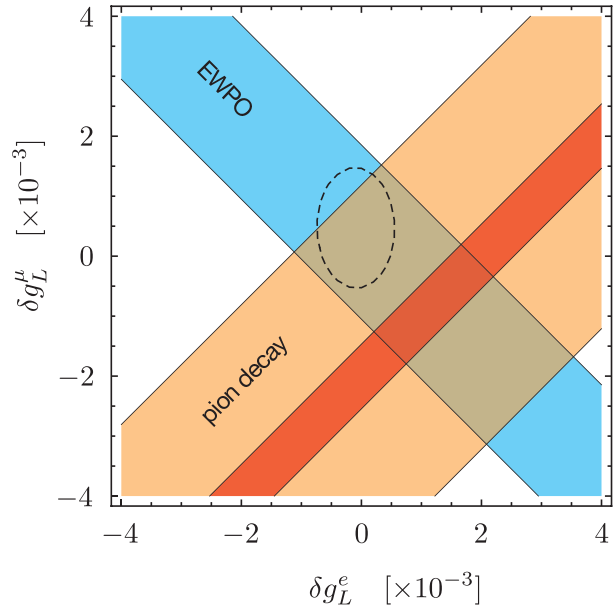}  
 \end{center}
 \caption{Constraint and sensitivity of the contribution to the lepton non-universal coupling to the $W$ (and $Z$) boson. 
 The axes are $\delta g^\ell_L \equiv \delta g^{W \ell \nu}$ with $\ell = e, \mu$.
 The orange and blue regions are currently allowed at 95\% CL by $R_{\pi \to e/\mu }$ and EWPOs, respectively. 
 The red region is a future sensitivity from $R_{\pi \to e/\mu }$ (Eq.~\eqref{eq:senst2}), where the central value is chosen to be the same as the current result (Eq.~\eqref{eq:senst1}). 
 If $\delta g^{Z \ell\ell}_L = \delta g^{W \ell \nu _{\ell }} \equiv \delta g^\ell_L$ is satisfied by the new physics model, EWPOs provide a more tight limit. 
 The region in the circle with the black dashed line is allowed at 95\% CL.}
 \label{fig:constraint}
\end{figure}

In this section, current bounds and future prospects of the lepton universality are discussed. 
Those of the pion decay are obtained in Eqs.~\eqref{eq:decaypi1} and \eqref{eq:decaypi2}. 
In Fig.~\ref{fig:constraint}, the current bound and future sensitivity are displayed by the orange and red regions, respectively. 
The axes are $\delta g^\ell_L \equiv \delta g^{W \ell \nu}$ with $\ell = e, \mu$.
The SM prediction corresponds to $\delta g^\ell_L = 0$.
Here, 95\% CL regions are shown.
The central value of the future sensitivity is set to be that of the current result \eqref{eq:senst1}. 
It is found that, for a given $\delta g^{We\nu _e}$, the sensitivity is expected to be improved by a factor of 5 or more in the future.

On the other hand, the current limit at 95\% CL from EWPOs (Sec.~\ref{sec:ewpo}) is shown by the blue region. 
Here, only $\delta g^{W \ell \nu}$ is considered, and $g^{Z\ell\ell}_L$ is assumed to be zero.
Thus, the contributions to $G_F$ \eqref{eq:corrGF} and $\Gamma_W$ \eqref{eq:corrGW} are taken into account.
The limit is mainly determined by the former. 
The uncertainty is smaller than the current result of the pion decay (also those from $K$ and $\tau$).
However, it is found that the pion decay can have a better sensitivity to probe the lepton universality in the future, as observed by comparing the red and blue bands. 

Furthermore, it is stressed that, in the figure, the slope of the band is different between the pion decay and EWPOs. 
The former depends on $\delta g^{W \mu \nu  _{\mu }} - \delta g^{We\nu _e}$, while the latter on $\delta g^{W \mu \nu  _{\mu }} + \delta g^{We\nu _e}$. 
Obviously, the couplings cannot be determined separately if we use only one measurement. 
However, if we combine these two results, it is possible to determine $\delta g^{W \mu \nu  _{\mu }}$ and $\delta g^{We\nu _e}$ individually without assuming new physics models. 
The corrections are restricted at 95\% CL in the range:
\begin{align}
-1.1 \times 10^{-3} < \delta g^{We\nu _e} < 3.5 \times 10^{-3}, ~~~~~
-3.1 \times 10^{-3} < \delta g^{W \mu \nu _{\mu }} < 1.5 \times 10^{-3}.
\end{align}
The error is dominated by the experimental uncertainty of the pion decay.
In future, the pion decay rate will be measured accurately.
The error is expected to be reduced by a factor of 2, where the experimental uncertainty of EWPOs dominates the error. 
Hence, the pion decay can play a complementary role in determining the contributions to the electron and muon interactions with the $W$ boson.

Finally, let us consider a case when $\delta g^{Z \ell\ell}_L = \delta g^{W \ell \nu _{\ell }} \equiv \delta g^\ell_L$ is satisfied.
This corresponds to the extra lepton model in the next section. 
When the lepton coupling to the $Z$ boson is tightly correlated with that to the $W$ boson, EWPT constrains $g^{Z ee}_L$ and $g^{Z \mu\mu}_L$ individually.
In Fig.~\ref{fig:constraint}, the 95\% CL limit from EWPOs is shown by the circle with the black dashed line. 
Since the experimental result of the $Z \to e^+ e^-$ decay is accurate, the constraint on $ \delta g^{W e \nu _e }$ is tightened compared to the blue region. 
However, the EWPT accuracy of $\delta g^{Z \mu\mu}_L$ is comparable to that of $\delta g^{W \mu  \nu _{\mu } }$, and thus, the limit does not change so much for the muon couplings. 
It is found that the EWPT limit on the lepton non-universal coupling with the $W$ and $Z$ boson is currently more severe than the bound from the pion decay. 
However, the lepton non-universal coupling, especially of the muon, can be probed more sensitively by the pion decay in the near future.

\vspace*{5mm}


\section{Extra Lepton Model} \label{sec:model}

Let us apply the result in the previous section to an extra lepton model~\cite{Kannike:2011ng,Dermisek:2013gta}. 
We introduce SU(2)$_L$ vector-like doublets $(L_L \oplus L_R)$ and singlets $(E_L \oplus E_R)$, which are coupled only to the muon.
It is assumed that the $(L_L \oplus L_R)$ and $(E_L \oplus E_R)$ have the same quantum numbers as the left- and right-handed SM leptons, respectively. 
Their interaction and mass terms are given as
\begin{align}
 \mathcal{L} =  
 & 
 - \bar{\ell}_{L} Y_{\mu } \ell_{R} H 
 - \bar{\ell}_{L} \lambda _E E_R H 
 - \bar{L}_L \lambda _L \ell_{R} H 
 - \lambda  \bar{L}_L E_R H 
 - \bar{\lambda } H^{\dagger }\bar{E}_L L_R \notag \\
 &
 - M_L \bar{L}_L L_R 
 - M_E \bar{E}_L E_R + \rm{h.c.}, 
\end{align}
where $\ell$ denotes the SM lepton, and $H$ is the Higgs boson. 
The terms in the first line represent the Yukawa interactions, while those in the second line are vector-like mass terms. 
The components of the SU(2)$_L$ doublets and singlets are represented as
\begin{equation}
 \ell _{L} = \begin{pmatrix} 
                  \nu _{\mu } \\
                  \mu _{L} \\
             \end{pmatrix},~~~
 \ell_R = \mu _R,~~~
 L_{L,R} =  \begin{pmatrix} 
                  L^0 _{L,R} \\
                  L^- _{L,R} \\
             \end{pmatrix},~~~
 H = \begin{pmatrix}
            0 \\
            v + \frac{h}{\sqrt{2}} \\
       \end{pmatrix},
\end{equation}
where $v \simeq 174\,{\rm GeV}$ is the vacuum expectation value of the Higgs field. 
After the electroweak symmetry is broken, the mass term of the charged leptons becomes
\begin{equation}
 -\mathcal{L}_{\rm mass} = 
 (\bar{\mu }_{L}, \bar{L}^-_{L},\bar{E}_L) \mathcal{M}_{\mu } \begin{pmatrix} \mu _{R} \\ L^-_{R} \\ E_R  \\ \end{pmatrix} + \rm{h.c.},
\end{equation}
where $\mathcal{M}_{\mu}$ is the $3 \times 3$ mass matrix for the muon and the vector-like leptons,
\begin{equation}
 \mathcal{M}_{\mu} = 
 \begin{pmatrix} 
    Y_{\mu } v & 0 & \lambda _E v \\
    \lambda _L v & M_L & \lambda v \\
    0 & \bar{\lambda } v & M_E \\
 \end{pmatrix}.
 \label{eq:modelY}
\end{equation}
This matrix is diagonalized by unitary matrices $U_{L,R}$ as
\begin{equation}
  U_L ^{\dagger} \mathcal{M}_{\mu} U_R = \text{diag}(m_{\mu }, m_{\ell_2} ,m_{\ell_3}).
  \label{eq:mixmat}
\end{equation}
Here, $m_{\mu }$ is the muon mass, and $m_{\ell_{2,3}}$ are the heavy lepton masses in the mass eigenstate basis.

In this letter, we study the lepton non-universal coupling to the $W$ boson and the left-handed one to the $Z$ boson. 
They are determined by $U_L$.
In the limit of $\lambda _E v$, $\lambda _L v$, $\bar{\lambda } v$, $\lambda v  \ll M_E$, $M_L$, it is approximately written as
\begin{align}
 &U_L = 
 \begin{pmatrix}
     1 - v^2 \frac{\lambda _E^2}{2M_E^2} 
   & - v^2 
     \left( 
       \frac{\lambda _E}{M_L} 
       \frac{\bar{\lambda } M_E + \lambda M_L}{M_E^2 - M_L^2 } - 
       \frac{Y_{\mu }\lambda _L}{M_L^2}
     \right)
   & v \frac{\lambda _E}{M_E} \\
     - v^2 \frac{\bar{\lambda } \lambda _E M_L - Y_{\mu } \lambda _L M_E}{M_L^2 M_E } 
   & 1 - v^2 \frac{(\lambda M_E + \bar{\lambda } M_L )^2}{2(M_E^2 - M_L^2 )^2}
   & v \frac{\lambda M_E + \bar{\lambda } M_L }{M_E^2 - M_L^2} \\
     - v \frac{\lambda _E}{M_E}
   & - v \frac{\lambda M_E + \bar{\lambda } M_L }{M_E^2 - M_L^2} 
   & 1 - v^2 \frac{\lambda _E^2}{2M_E^2} 
       - v^2 \frac{(\lambda M_E + \bar{\lambda } M_L )^2}{2(M_E^2 - M_L^2 )^2} \\
 \end{pmatrix}.
 \label{eq:appmat}
\end{align}
After the diagonalization, the muon couplings to the $W$ and $Z$ bosons are modified.\footnote{The Higgs coupling is also modified from the SM prediction~\cite{Kannike:2011ng}.}
The corrections become
\begin{align}
\delta g^{W \mu \nu _{\mu }} 
&= - \left[ (U_L)_{11} -1 \right]
\simeq \frac{1}{2} \left( \frac{\lambda _E v}{M_E} \right) ^2, \notag \\
\delta g^{Z \mu \mu}_L 
&= \frac{1}{2} (U^{\dagger }_L)_{13}(U_L)_{31} 
\simeq \frac{1}{2} \left( \frac{\lambda _E v}{M_E} \right) ^2.
\label{eq:corrEL}
\end{align}
It is noticed that both are determined by $(\lambda _E v / M_E)^2$, and 
$\delta g^{W \mu \nu _{\mu}} = \delta g^{Z\mu\mu}_L$ is satisfied approximately.\footnote
{
The right-handed coupling $\delta g^{Z\mu\mu}_R$ also receives a correction, $\sim (\lambda_L v/M_L )^2$. 
It is suppressed when $M_L$ is larger or $\lambda_L$ is small.
We neglects its contribution to EWPOs, for simplicity.
}
On the other hand, there is no correction to the electron coupling with the SM bosons, since the extra leptons do not couple to the electron in Eq.~\eqref{eq:modelY}.

The result in the previous section is applied to Eq.~\eqref{eq:corrEL}. 
From Fig.~\ref{fig:constraint}, the constraint on $(\lambda _E v / M_E)^2$ is derived.
Currently, the most tight limit is imposed by EWPT. 
The bound is estimated at 95\% CL as\footnote
{
The contribution to the muon $g-2$ depends also on other parameters in the extra lepton models.
See Refs.~\cite{Kannike:2011ng,Dermisek:2013gta} for the explicit formula. 
}
\begin{align}
  - 1 \times 10^{-3} \lesssim 
  \left( \lambda _E v / M_E \right )^2 \lesssim 
  3 \times 10^{-3}.
\end{align}
The result is consistent with zero, i.e., the SM prediction.
This is stronger than the limit from the pion decay (also from the $K$ and $\tau$ decays). 
In the near future, the pion decay can have a better sensitivity to the coupling. 
The error of $(\lambda _E v / M_E)^2$ will be reduced to be about $1 \times 10^{-3}$ at 95\% CL, which is better by a factor of 2 than EWPT. 
Thus, the search for the lepton non-universality with the leptonic pion decay can compete with the limit with EWPT. 

Finally, let us touch on other extra lepton models.
If the contribution to the muon coupling to the $Z$ boson is suppressed with keeping those to the $W$ boson sizable, the pion decay not only has a better sensitivity, but also provides a complimentary information in determining the corrections to the $W\ell\nu$ coupling, as mentioned in the previous section.


\section{Conclusion}

In this letter, we discussed the lepton flavor non-universal coupling to the $W$ and $Z$ bosons. 
Here, the left-handed lepton interactions are considered. 
The electron/muon coupling to the $W$ boson is currently constrained by EWPT.
We showed that the ratio of the leptonic pion decay, $R_{\pi \to e/\mu }$, is expected to be more sensitive to the non-universality in the near future.
Furthermore, it was found that the ratio is complimentary to determine the electron and muon non-universal couplings individually without assuming new physics models.

We applied the result to an exotic lepton model. 
In the model, the left-handed muon coupling with the $Z$ boson is tightly correlated with that with the $W$ boson, and the EWPT limit is severe.
It was discussed that the pion decay will be able to compete with or be better than the EWPT result.
Thus, the measurement of the leptonic pion decay is expected to play an important role in searching for effects of the exotic lepton models.

\section*{Acknowledgements}
This work was supported by JSPS KAKENHI Grant No.~23740172 (M.E.) and 25-10486 (T.Y.).
The work of T.Y. is supported by a JSPS Research Fellowship for Young Scientists 
and by an Advanced Leading Graduate Course for Photon Science grant.


\providecommand{\href}[2]{#2}
\begingroup\raggedright

\endgroup

\end{document}